\documentclass[twocolumn,prl]{revtex4}

\usepackage{times}
\usepackage{epsfig}
\usepackage{amsmath,amssymb}

\begin{document}

\title{Non-Markovian dynamics of clusters during nucleation}

\author{J.\ Kuipers}
\email{jkuipers@phys.uu.nl}
\affiliation{Institute for Theoretical Physics, Utrecht University, Utrecht, The Netherlands}
\author{G.\ T.\ Barkema}
\affiliation{Institute for Theoretical Physics, Utrecht University, Utrecht, The Netherlands}
\affiliation{Instituut-Lorentz for Theoretical Physics, Leiden University, Leiden, The Netherlands}

\begin{abstract}
Most theories of homogeneous nucleation are based on a
Fokker-Planck-like description of the behavior of the mass of
clusters.  Here we will show that these approaches are incomplete for
a large class of nucleating systems, as they assume the effective
dynamics of the clusters to be Markovian, i.e., memoryless. We
characterize these non-Markovian dynamics and show how this influences
the dynamics of clusters during nucleation. Our results are validated
by simulations of a three-dimensional Ising model with locally
conserved magnetization.
\end{abstract}

\maketitle

Nucleation is the process where a stable nucleus spontaneously emerges
in a metastable environment. Excellent books and reviews exist on this
topic; in his recent book, Kashchiev~\cite{kashchiev} lists about 30
books and 40 review articles on nucleation.  Examples of nucleation
abound, for instance the formation of droplets in undercooled gasses
and of crystals in undercooled liquids. The process is thermally
activated and is key to understanding various subjects in biophysics,
polymer physics, and chemistry. The physics behind it has long been
studied and the simplest version is known as {\it classical nucleation
theory} (CNT)~\cite{becker,cnt}.

In CNT the variations in the mass of a nucleus are described as a
Markovian stochastic process in which single units attach and detach
from the nucleus. The probability $p(m,t)$ that a nucleus has mass $m$
at time $t$ evolves via the Fokker-Planck equation~\cite{cnt,kampen}
\begin{equation}
\frac{\partial p(m,t)}{\partial t} = 
  \frac{\partial}{\partial m} \left[R(m) 
  \left(\beta\,\frac{\partial F}{\partial m}
  +\frac{\partial}{\partial m}\right)p(m,t) \right],
\end{equation}
with $R(m)$ the rate at which clusters of mass $m$ grow to clusters of
mass $m+1$, $F(m)$ the free energy of a cluster of mass $m$ and
$\beta$ the inverse temperature. Phenomenological expressions for
$R(m)$ and $F(m)$ are provided to complete the theory. $F(m)$
increases for small $m$, but decreases for large $m$, and it retains
its maximal value at the so-called {\it critical nucleus mass}
$m_c$. Therefore, clusters with mass below $m_c$ tend to shrink,
whereas clusters larger than $m_c$ tend to grow.
Starting at time $t=0$ with cluster mass $m(0)=m_c$, the mass
evolves diffusively in time: its mean square change, averaged
over all realizations,
$\langle\Delta m(t)^2\rangle \equiv \langle(m(t)-m_c)^2\rangle$, 
grows linearly with time for small deviations from the critical mass.

\begin{figure}
\centering
\includegraphics[width=\columnwidth]{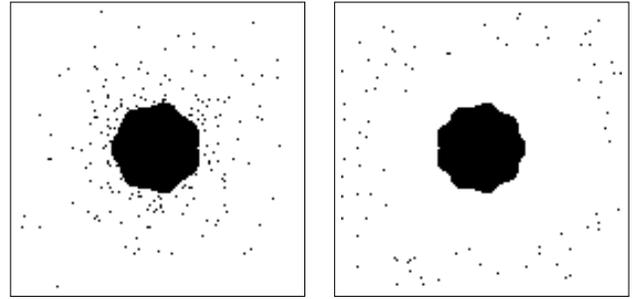}
\caption{Illustration of memory effects in the Ising model with local
  spin exchange dynamics (exaggerated). Although they have equal
  masses, the cluster in the left panel is likely to grow as it has
  just shrunk, hence has higher density surrounding it. In the right
  panel this is vice versa.}
\label{fig::growshrink}
\end{figure}

Although, qualitatively, there is ample experimental verification of
the predictions of CNT, quantitatively, the nucleation rates it
predicts may differ from experiment by 5 to 10 orders of
magnitude~\cite{auer}. CNT works really well if the nucleation process
is correctly described by a Markov process with units attaching and
detaching as uncorrelated events. This is more or less the case in,
for example, the Ising model with spin flip dynamics. However, in the
presence of a local conservation law, which, for example, is the case
in binary mixtures of fluids or gasses and the Ising model with local
spin exchange dynamics, CNT shows serious shortcomings due to the
neglect of memory effects by describing the cluster growth as a Markov
process. The most dominant of these usually is the strong correlation in
time between attachment and detachment events: after a particle detaches
from the nucleus, it remains in its neighborhood for a relatively
long time and is therefore likely to reattach. In fact, the theory
of Brownian motion~\cite{brownian} states that in two dimensions,
\emph{every} detached particle will eventually return to the cluster,
possibly after an extremely long time, and that in three dimensions, a fraction
of the detached particles will never return.
Figure \ref{fig::growshrink} illustrates this memory effect in the Ising
model with local spin exchange dynamics. Two clusters of equal mass
are shown. The cluster in the left panel has just shrunk, consequently
has a higher density of particles surrounding it, which enhances the
probability that the cluster will grow; in the right panel, the
opposite is happening and the cluster is more likely to shrink. These
memory effects result in behavior quite different from CNT. The
relation between the driving force $-\partial F/\partial m$ and the drift velocity
is more complicated and the diffusion is anomalous, i.e.\
the mean square displacement scales non-linearly with time.

We develop a theory for the variation in the cluster mass, where these
memory effects are taken into account and show that on different time
scales different types of dynamical behavior take place. We
concentrate on states in three-dimensional systems with a small
gradient in free energy, which is the case for, for instance,
near-critical nuclei. In that case, three distinct regimes show up. On
very short time scales the mean square growth of a cluster scales
linearly with time like in an ordinary diffusive process. This regime
is followed by a regime of anomalous diffusion with exponent of one
half (i.e.\ $\left<\Delta m(t)^2\right>\sim t^{1/2}$) and finally on
large time scales linear growth occurs again, but with a much smaller
prefactor than in the first regime. We validate our theory with
extensive simulation results on the mean square growth of
near-critical nuclei in a three-dimensional Ising model with local
spin exchange dynamics.

\begin{figure}
\centering
\includegraphics[width=0.9\columnwidth]{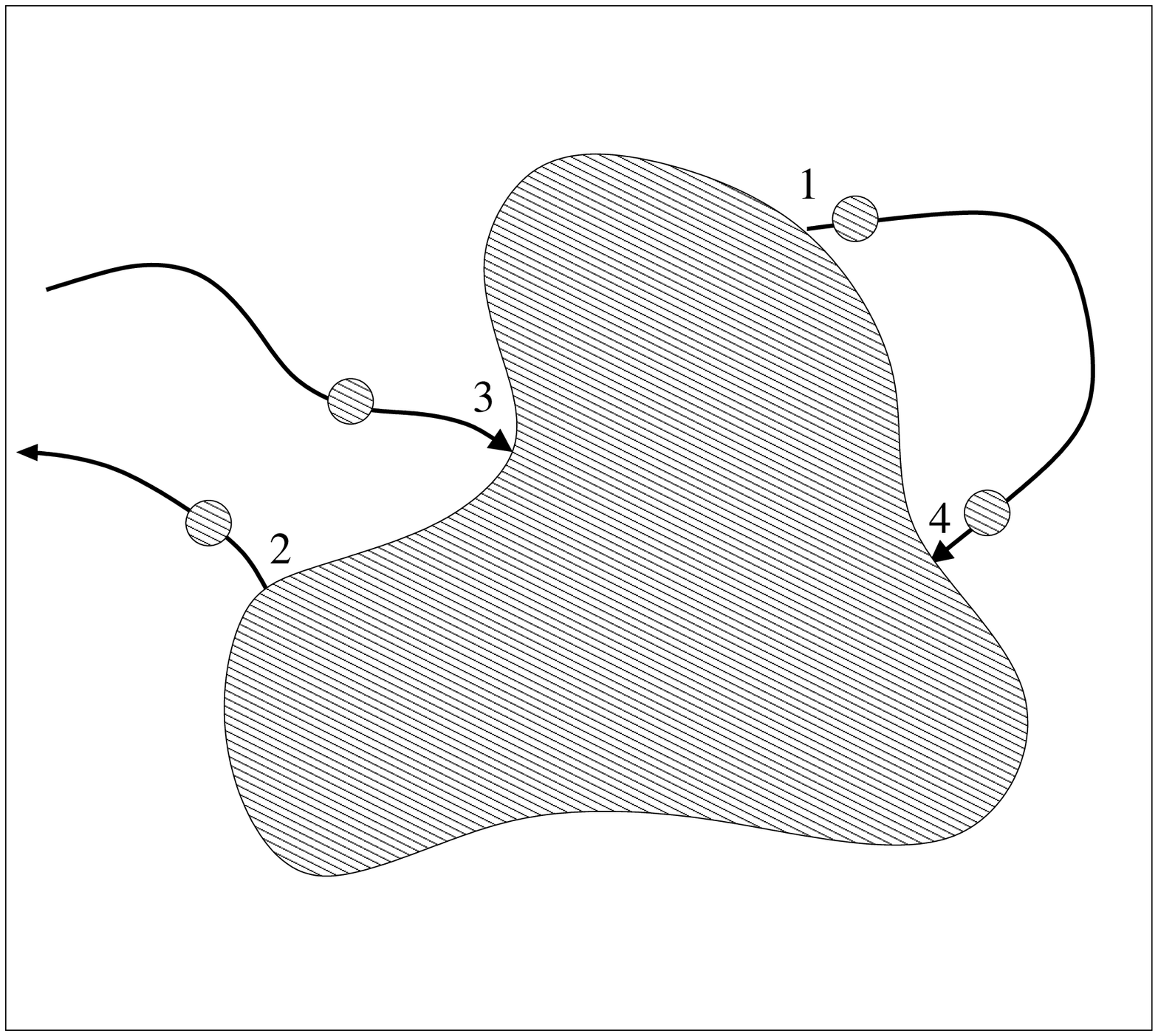}
\caption{The four effects causing variations in the cluster mass: 1)
the emission of returning particles, 2) the emission the non-returning
particles, 3) the absorption of non-returning particles, and 4) the
absorption of returning particles.}
\label{fig::foureffects}
\end{figure}

In our model, we consider a cluster in a dilute environment (like in
figure \ref{fig::growshrink}). The variations in the cluster mass are
caused by four effects, namely 1) the emission of particles from the
cluster, that are returning (with return probability $p_r$), 2) the
emission of non-returning particles, 3) the absorption of particles
from far away (i.e.\ not returning from previous emission), and 4) the
absorption of particles that are returning. These four effects are
illustrated in figure
\ref{fig::foureffects}. The first three effects are independent and
are described by random functions $\xi_r(t)$, $\xi^-_{nr}(t)$ and
$\xi^+_{nr}(t)$, respectively. Each of these random functions consists
of a series of delta functions at Poisson distributed random times, so
that the cluster mass $m(t)$ is integer at all times. Their average
values obey
\begin{equation}
\langle\xi_{nr}^+(t)\rangle - \langle\xi_{nr}^-(t)\rangle = v(m(t)),
\label{eqn::determine1}
\end{equation}
with $v(m)$ the systematic growth of a cluster of mass $m$. The
variations
$\delta\xi_\alpha(t)\equiv\xi_\alpha(t)-\langle\xi_\alpha(t)\rangle$
obey
\begin{equation}
\sum_\pm \langle\delta\xi^\pm_{nr}(t)\delta\xi^\pm_{nr}(t')\rangle
 = 2 (1-p_r) D_M(m(t)) \delta(t-t'),
\label{eqn::determine2}
\end{equation}
and
\begin{equation}
\langle\delta\xi_r(t)\delta\xi_r(t')\rangle = p_r D_M(m(t)) \delta(t-t'),
\label{eqn::determine3}
\end{equation}
with $D_M(m)$ the diffusion coefficient describing the short time mass
variations of a cluster of mass $m$. Equations
(\ref{eqn::determine1})-(\ref{eqn::determine3}) fully determine the
three random functions $\xi_\alpha(t)$.

The absorption of returning particles is correlated to their emission:
a returning particle emitted at time $\tau$ is returning at time
$\tau+T_\tau$, with $T_\tau$ described by a return time distribution $\mu$.
Putting this together, one may
describe the evolution of a cluster by the stochastic differential
equation:
\begin{align}
\dot{m}(t) & = 
  \xi_{nr}^+(t) - \xi_{nr}^-(t) - \xi_r(t) \nonumber \\
  & \qquad + \int_{-\infty}^t\!\!\!d\tau\,\delta(t-\tau-T_\tau)\,\xi_r(\tau).
\label{eqn::dotn}
\end{align}
The first two terms are basically CNT and the addition of the other
two terms is new. This stochastic differential equation is too hard to
be solved in general and therefore we restrict ourselves first to
near-critical clusters. On average critical clusters absorb equally
many particles as they emit, so we assume $v(m)=0$. Furthermore, we
assume that the diffusion coefficient is $m$-independent:
$D_M(m)=D_M$. The average growth $\langle \Delta m(t)\rangle$ is then
vanishing and the mean square growth can be calculated from
\begin{equation}
\left<\Delta m(t)^2\right>
\;=\; \int_0^t\!\!\!d\tau \int_0^t\!\!\!d\tau' \left<\dot{m}(\tau)\dot{m}(\tau')\right>,
\end{equation}
which, after substituting equation (\ref{eqn::dotn}) twice and using
(\ref{eqn::determine2}) and (\ref{eqn::determine3}), results in the
following expression
\begin{equation}
\left<\Delta m(t)^2\right> 
\;=\; 2 D_M\left(t - p_r \int_0^t\!\!\!d\tau(t-\tau)\mu(\tau)\right).
\label{eqn::nucleation}
\end{equation}
The first term in this equation is ordinary diffusion, which is
suppressed by the last term, since emission and absorption are
correlated at larger time scales.

Next we investigate various limits of the equation. We assume that the
time scale of spontaneous fluctuations (i.e.\ $t\approx1/D_M$) is much
smaller than the time scale on which particles are typically
returning. For small times the leading term in equation
(\ref{eqn::nucleation}) results in
\begin{equation}
\left<\Delta m(t)^2\right> \;\approx\; 2D_Mt.
\end{equation}
At these time scales attachments and detachments occur at different
places on the cluster's surface as independent events, and therefore
the mean square cluster growth scales linearly with time.

To investigate $\langle \Delta m(t)^2\rangle$ at larger times we have
to specify $\mu(t)$ in more detail. We assume that a detached particle
makes a three-dimensional random walk with diffusion coefficient
$D_B$. Furthermore, it starts at a distance $\delta R$ separated from
the cluster, which we consider to be a sphere of radius $R$. This
gives for the return probability $p_r=\frac{R}{R+\delta R}$ and for
the return time distribution~\cite{hitsphere}
\begin{equation}
\mu(t) \;=\; 
  \frac{\delta R\,\exp\left(-\delta R^2/4D_Bt\right)}{\sqrt{4\pi D_B}\,t^{3/2}}.
\label{eqn::mu3D}
\end{equation}
Asymptotically at large times $\mu(t)\sim t^{-3/2}$, but it is cut off
at small times by the exponential. Using this distribution to
calculate the mean square cluster growth for larger times results in
\begin{equation}
\left<\Delta m(t)^2\right> 
\;\approx\; 2(1-p_r)D_Mt + \text{const}\;\sqrt{t}.
\label{eqn::largetimes}
\end{equation}
At large time scales, the first term dominates, and the cluster
dynamics is determined by particles being emitted to and absorbed from
far away with rate $2(1-p_r)D_M$. At those large times, the change in
cluster size is determined by non-returning absorbed and emitted
particles, and those particles can be treated as independent
events. At intermediate time scales, however, the anomalous diffusion
dominates if $p_r$ is close to one, which it typically is. This
behavior is caused by emitted particles returning, much comparable to
single file diffusion~\cite{singlefile}. Note that if the return
probability is one, the anomalous diffusion lasts forever. This may
happen, for example, in many two-dimensional systems, although the
long time behavior is different from equation (\ref{eqn::largetimes}),
due to a different return time distribution $\mu(t)$.

\begin{figure}
\centering
\includegraphics[width=\columnwidth]{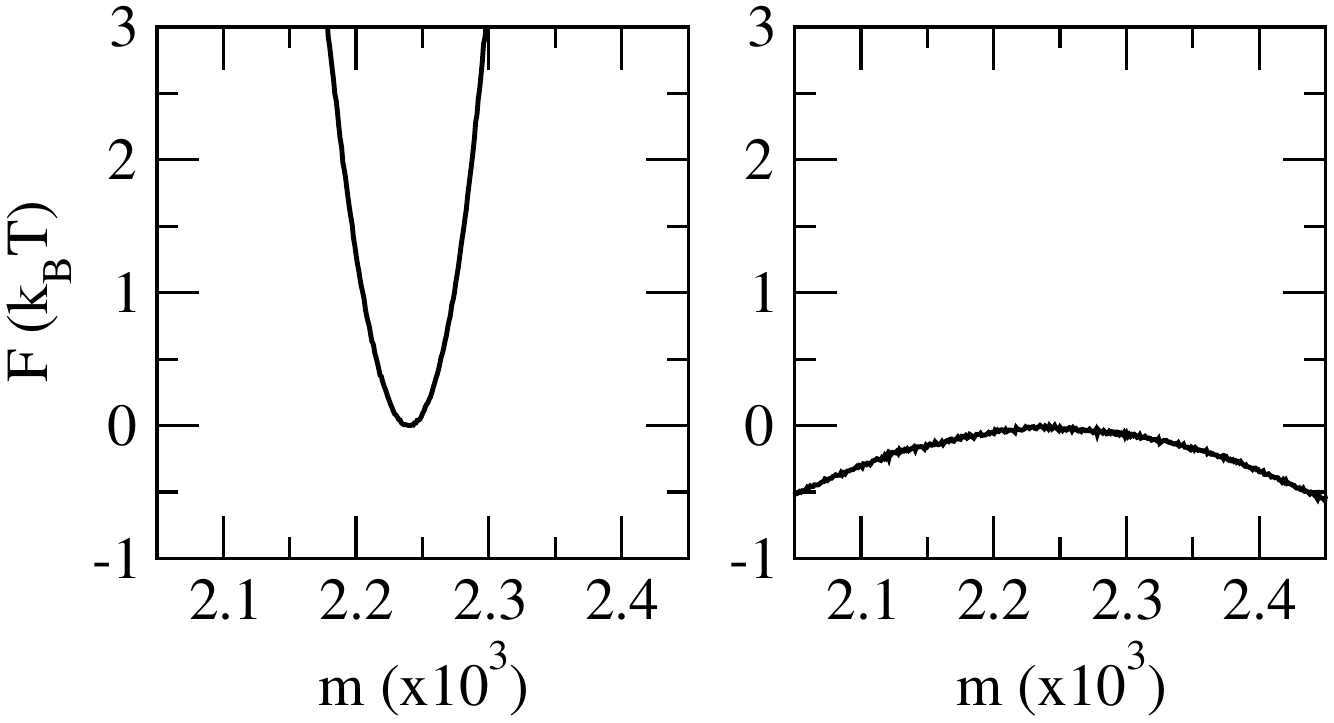}
\caption{Free energy landscapes of a finite-sized Ising model, as a function of
cluster mass; in the
left panel with conserved magnetization and in the right panel
without. The density in the left panel and the external magnetization
in the right panel are tuned, such that the extrema of the potentials
coincide.}
\label{fig::free}
\end{figure}

To validate the presented theoretical picture, simulations were
performed on a three-dimensional Ising model with local spin exchange
(Kawasaki) dynamics with Metropolis acceptance
probabilities~\cite{kawasaki,MCbook}, a prototypic system to study
nucleation. The Hamiltonian is given by
\begin{equation}
H \;=\; -J \sum_{\langle i,j\rangle} s_i s_j + h \sum_i s_i,
\end{equation}
with the first sum over all pairs of adjacent sites. A
three-dimensional cubic lattice of size $32\times32\times32$ with
periodic boundary conditions has been used for the simulations. The
temperature is chosen as $k_B T=2.5J$, well below the critical
temperature of $k_BT_c\approx 4.5J$~\cite{MCbook}. An oversaturated
initial configuration is first brought into equilibrium under constant
magnetization, resulting in a single large cluster of $m_c$ spins in
equilibrium with its surroundings with a density of $1.2$ percent.
Next, besides the spin exchange moves, we also perform spin flips in
three strips of the box (one in each principal direction), as far away
from the cluster as possible. These spin flip moves mimic an infinite
reservoir of up-pointing spins, and render the cluster instable. The
strength of the external field $h$ is fixed at the value giving the
initial cluster a critical size, hence zero average growth rate. The
free energy as a function of cluster mass is plotted in figure
\ref{fig::free}, before (left) and after (right) lifting the
constraint on the magnetization.

\begin{figure}[b]
\centering
\includegraphics[width=\columnwidth]{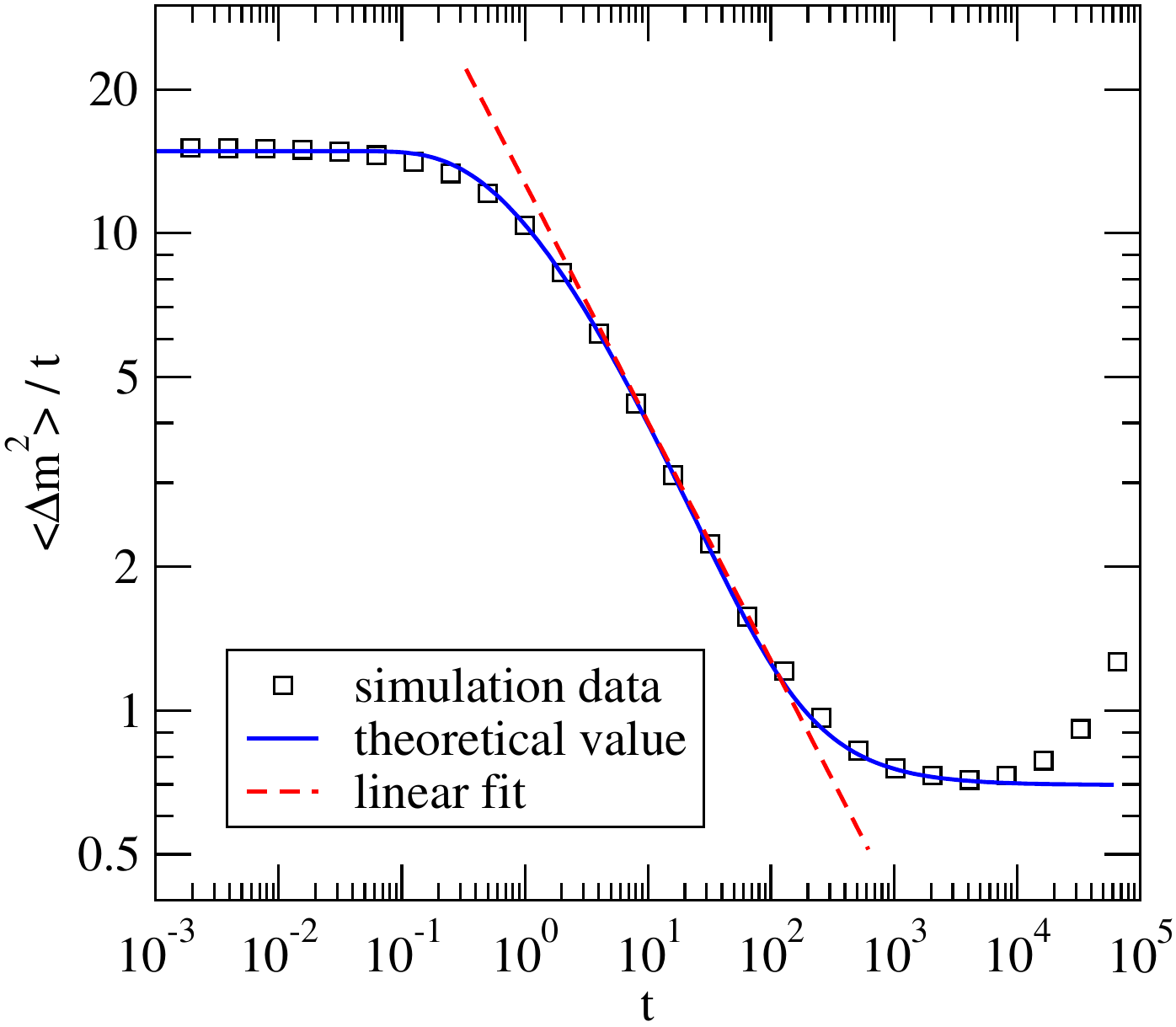}
\caption{Measurements of $\langle\Delta m(t)^2\rangle/t$ in the Ising
  model, plotted together with the theoretical value of equation
  (\ref{eqn::nucleation}) and a linear fit with a slope of minus one
  half. A zero slope indicates ordinary diffusion.}
\label{fig::results}
\end{figure}

The time evolution of this cluster is then measured and the resulting
mean square change in cluster mass as a function of time, averaged
over about 70,000 realizations, is shown in figure \ref{fig::results}.
For comparison, our theoretical estimate of $\langle \Delta
m(t)^2\rangle$ is plotted with the simulation data. Our theory
requires as input the function $\mu(t)$, with its parameters $D_B$,
$R$ and $\delta R$, and the parameters $D_M$ and $p_r$. They are
chosen as follows. The distribution $\mu(t)$ is as in equation
(\ref{eqn::mu3D}), multiplied by an exponential, since the system has
a finite size and long return times are exponentially surpressed. The
result depends very little on the chosen exponent. Furthermore,
$D_B=\textstyle\frac12$, in accordance with the definition of time in
our model, and $R$ is chosen such that $m_c=\textstyle\frac43\pi R^3$.
The parameter $\delta R$ is the initial distance between the cluster
and a spin which has just detached from it. If the spin detaches
radially, this distance equals the lattice spacing, but the effective
distance can be smaller by a factor of up to two in other directions.
We chose $\delta R=\frac{8+\pi^2}{8\pi}$, since that is the average
distance between a spherical cluster and the sites neighboring it on
the square lattice. The diffusion coefficient $D_M$ in equation
(\ref{eqn::nucleation}) and subsequently the return probability $p_r$
are fitted.

The theory captures the general trend well. For short times ordinary
diffusion is observed in the simulation data, after which anomalous
diffusion with exponent one half is clearly present, as indicated by
the dashed line in figure \ref{fig::results}. For large times,
however, the cluster is so far out of equilibrium that the gradient in
the free energy is non-zero, so that $\langle \Delta m(t)\rangle\neq0$
and $\langle \Delta m(t)^2\rangle$ grows super-linearly with time. We
verified that the same results hold, qualitatively, for different
values of the temperature.

Next, we turn to the consequences of these memory effects on
nucleation times. To obtain nucleation times, equation (\ref{eqn::dotn})
should be solved in the presence of a free energy barrier, which
results in a systematic growth $v(m)$ in equation
(\ref{eqn::determine1}). The phenomenological expression for the free
energy from CNT~\cite{becker,cnt} or quadratic approximation could be
used. This work is still in progress and therefore we resort to
scaling arguments in the meantime.

Instead of the nucleation time, we focus on the vaporization time
(i.e.\ the time it takes for a critical nucleus to vaporize); these
two times are connected via detailed balance by $T_\text{nucl} \approx
e^{\beta\Delta F} T_\text{vap}$, hence $T_\text{vap}$ is the
pre-exponential factor of the nucleation time. This vaporization time
is mainly dictated by the time a critical cluster resides in a region
near the top of the free energy barrier; after that it vaporizes
relatively fast due to the gradient in the free energy. We define this
region near the top as the region where the free energy is less than
$k_BT$ below the maximum value $F(m_c)$ and call its width $\delta
m$. In CNT the residence time for this region, often referred to as
the Zeldovich factor~\cite{zeldovich}, scales as
$T_\text{res}\sim\delta m^2$. If, in equation (\ref{eqn::largetimes}),
the anomalous diffusion is taken over by the normal diffusion at the
residence time, the memory effects only result in rescaling the
diffusion coefficient by a factor of $1-p_r$. However, if the cluster
dynamics show sub-diffusive behavior up to this residence time, then
$\langle\Delta m(t)^2\rangle\sim\sqrt{t}$, so that the residence time
is proportional to $\delta m^4$. Which behavior occurs depends on
numerous variables, such as the shape of the free energy landscape and
the mobility of detached particles.

Another view on the consequences of memory effects on the nucleation
time is that, compared to CNT with rescaled diffusion coefficient
$D_\text{eff}=(1-p_r)D_M$, anomalous
diffusion is present in addition to CNT's ordinary diffusion. This
results in increased mass fluctuations around the critical mass, hence
decreases the residence time and therefore also the nucleation time. 

In summary, we have demonstrated that memory effects are playing an
important role in the dynamics of nucleation. The time evolution of
nucleating clusters shows consequently anomalous diffusion. These
memory effects can be captured by a simple stochastic differential
equation~(\ref{eqn::dotn}), which gives measurable results for
critical clusters. These effects probably also have impact on the
nucleation times. Analytic treatment of this equation in the presence
of a free energy barrier is therefore an interesting topic for future
research and might lead to new quantitative predictions for nucleation
times.

We thank Henk van Beijeren for useful discussion.

\end{document}